\documentclass[preprintnumbers,amsmath,amssymb,floatfix,superscriptaddress,nofootinbib]{revtex4}
\pdfoutput=1
\usepackage{amsmath,hyperref,amssymb}
\usepackage{graphicx}
\usepackage{dcolumn}
\usepackage{bm}
\usepackage{verbatim}
\usepackage{tabularx}
\usepackage{slashed}
\usepackage{cancel}
\usepackage[boxsize=2em,aligntableaux=center]{ytableau}
\usepackage{float}

\def\be{\begin{equation}}
\def\ee{\end{equation}}
\def\bea{\begin{eqnarray}}
\def\eea{\end{eqnarray}}

\begin{document}

\vspace*{-2cm}

\title{Naturalness without new particles}
\author{Anson Hook}
\affiliation{Maryland Center for Fundamental Physics, University of Maryland, College Park, MD 20742, USA}


\begin{abstract}
\begin{centering}
{\bf Abstract}\\[4pt]

We demonstrate that the physics which resolves naturalness problems need not take the form of new particles and can sometimes manifest itself as higher dimensional operators.  As a proof of principle, we present a simple model where the scale of new particles is parametrically separated from that estimated via naturalness arguments applied to self-quartic couplings.  In this example, new particles appear far above the scale $m/\sqrt{\lambda}$, where $m$ is the mass of the particle and $\lambda$ is its self-quartic coupling.  The shift symmetry responsible for resolving the naturalness problem involves higher dimensional operators rather than new particles.  

\end{centering}

\end{abstract}

\vspace*{3cm}

\maketitle

\noindent\rule{\textwidth}{1pt}

\vspace*{1cm}

The Electroweak Hierarchy problem is one of the driving forces in particle physics (see e.g. Ref.~\cite{Dine:2015xga} for a review of the current state of affairs).  When the naturalness paradigm is applied to this problem, it is typically stated that new particles must appear by the TeV scale modulo accidental fine tunings.  In this article, we provide an interesting proof of principle that
the physics which resolves naturalness problems need not involve new particles.  

Consider a complex scalar $\Phi$ which is charged under a $\mathbb{Z}_N$ symmetry so that
\bea
\Phi \rightarrow e^\frac{2 \pi i}{N} \Phi .
\eea
The most general Lagrangian allowed by symmetries is
\bea
\label{Eq: one}
V = - f^2 | \Phi |^2 + | \Phi |^4 + \frac{\Phi^N}{\Lambda^{N-4}} + \cdots \, ,
\eea
where we take all coefficients to be $\mathcal{O}(1)$ numbers.  In the context of axions, this exact potential has been considered many times when discussing Planck suppressed corrections to the axion potential, see e.g. ~Ref.\cite{Kamionkowski:1992mf}.  
At the renormalizable level, there is an accidental $U(1)$ symmetry.  The leading-order operator that breaks this symmetry is shown above and is a higher-dimensional operator whose origin is unimportant.  Later, we will give a simple example where fermions with a mass $\gtrsim f$ can be integrated out to generate it.  
The radial mode is integrated out at an energy scale $f$.

At low energies there is a pseudo-Goldstone boson with the potential
\bea
V = - \epsilon^4 \cos \frac{N \phi}{f} \, , \qquad \epsilon^4 \sim \frac{f^N}{\Lambda^{N-4}} .
\eea
Expanding the potential, we see that the mass and quartic couplings of $\phi$ are 
\bea
\label{Eq: mass}
m_\phi^2 \sim \frac{\epsilon^4 N^2}{f^2} \, , \qquad \lambda \sim  \frac{\epsilon^4 N^4}{f^4} .
\eea
We will be concerned with an imaginary experimentalist who has discovered the pseudo-Goldstone boson and measured its mass and quartic coupling.
Said person would apply the standard naturalness arguments and expect the scale of new particles to be
\bea
\Lambda \sim \frac{m}{\sqrt{\lambda}} \sim \frac{f}{N} .
\eea
However, note that there are new particles, namely the radial mode, at $f$ rather than $f/N$.   Thus by taking $N$ large, we have a proof of principle that the mass scale of new particles can be parametrically separated from where naturalness arguments typically expect them to be.  If higher harmonics were generated instead, e.g. $V \sim \cos^m \left ( N \phi/f \right)$, the expectation would be that new particles are at the scale $f/\left ( N \sqrt{m} \right )$ further separating the scale where new physics is expected and the scale at which it actually appears.

The reason why the IR experimentalist incorrectly estimated the scale of new physics is because measuring the renormalizable Lagrangian was not enough information in order to determine the symmetries of the theory.  If the IR experimentalist proceeded to look for higher dimensional operators, he/she would eventually discover the higher dimensional operator
\bea
\frac{\epsilon^4 N^6 \phi^6}{f^6} \sim \frac{\lambda^2 \phi^6}{m^2}
\eea
It is important to note that this operator becomes strongly coupled at a scale $\gg f$ so that it would be difficult to find experimentally.  Upon measuring this operator, the experimentalist would note that this dimension six operator is algebraically related to the quartic coupling and mass terms and would suspect that there is a shift symmetry in the theory.  The useful operators of the theory are not $\phi^2$, $\phi^4$ and $\phi^6$ but instead only a single operator in the theory has been discovered, $\cos \frac{N \phi}{f}$.  As the coefficient of this operator, $\epsilon^4$, goes to zero there is an enhanced full shift symmetry for $\phi$.  It is thus technically natural to set $\epsilon^4$ small and there is no hierarchy/naturalness problem.  The dimensionful scale associated with the symmetry, $f/N$, is independent of the scale of new particles and thus cannot be used to predict the appearance of said particles.

At this point, we give a quick demonstration of how the higher-dimensional operator in Eq.~\ref{Eq: one} can be generated.  There are many ways to generate this operator, e.g. from theory space~\cite{ArkaniHamed:2001ca} or from discrete symmetries~\cite{Hook:2018jle}.  We will follow the discrete symmetry approach of adding to the theory a set of N vector-like fermions $\Psi_j$ and $\Psi_j^c$, which transform under a $\mathbb{Z}_N$ symmetry as
\bea
\Psi_j^{(c)} \rightarrow \Psi_{j+1}^{(c)} \, ,
\eea
with $j=0$ and $j=N$ identified.  Writing down the most general renormalizable Lagrangian, we have
\bea
\mathcal{L} \supset f^2 | \Phi |^2 - | \Phi |^4  + \sum_{j=1}^N \left ( m_f + y e^{2 \pi i j/N} \Phi \right) \Psi_j \Psi_j^c .
\eea
Taking $m_f > y f$, we can integrate out the fermions $\Psi$, which generates the higher-dimensional operator of interest ($y^N \Phi^N/m_f^{N-4}$) as well as some $U(1)$ symmetry preserving operators.

We now turn to the implications of this example for the naturalness argument.  We have shown that the physics which resolves naturalness problems could take the form of higher dimensional operators rather than new particles.  The natural question is to wonder if the solution to the Electroweak Hierarchy problem can also take the form of higher dimensional operators.  Currently we do not know whether or not this is possible.  However it is clear that the
moral of the story is that if such a solution were to exist, that one would search for higher dimensional operators rather than new particles.  Without an explicit example, it is not clear which higher dimensional operators to look for or what value to expect them to take (e.g. one might expect an operator of the form $\sim \lambda^2 \left (H H^\dagger \right )^3/m_H^2$).  The scale that suppresses these higher dimensional operators might be large, but one hopes that they are in reach of current experiments.

To conclude, we have demonstrated through an explicit counter-example that the solution to a naturalness problem need not involve new particles.  While this example has nothing to say about gauge and Yukawa quadratic divergences or about the cosmological constant, it would be interesting if similar models could be constructed for these other couplings.  This example suggests an interesting new approach to the naturalness paradigm.  Perhaps new physics will appear in the form of higher dimensional operators rather than new particles.

\section*{Acknowledgments}
We are very grateful to Zackaria Chacko, Junwu Huang, Matt Strassler, Raman Sundrum, and Gustavo Marques Tavares for useful discussions.  This research was supported in part by the NSF under Grant No. PHY-1620074 and by the Maryland Center for Fundamental Physics (MCFP).

\bibliography{ref.bib}

\begin{thebibliography}{4}
\expandafter\ifx\csname natexlab\endcsname\relax\def\natexlab#1{#1}\fi
\expandafter\ifx\csname bibnamefont\endcsname\relax
  \def\bibnamefont#1{#1}\fi
\expandafter\ifx\csname bibfnamefont\endcsname\relax
  \def\bibfnamefont#1{#1}\fi
\expandafter\ifx\csname citenamefont\endcsname\relax
  \def\citenamefont#1{#1}\fi
\expandafter\ifx\csname url\endcsname\relax
  \def\url#1{\texttt{#1}}\fi
\expandafter\ifx\csname urlprefix\endcsname\relax\def\urlprefix{URL }\fi
\providecommand{\bibinfo}[2]{#2}
\providecommand{\eprint}[2][]{\url{#2}}

\bibitem[{\citenamefont{Dine}(2015)}]{Dine:2015xga}
\bibinfo{author}{\bibfnamefont{M.}~\bibnamefont{Dine}}, \bibinfo{journal}{Ann.
  Rev. Nucl. Part. Sci.} \textbf{\bibinfo{volume}{65}}, \bibinfo{pages}{43}
  (\bibinfo{year}{2015}), \eprint{1501.01035}.

\bibitem[{\citenamefont{Kamionkowski and
  March-Russell}(1992)}]{Kamionkowski:1992mf}
\bibinfo{author}{\bibfnamefont{M.}~\bibnamefont{Kamionkowski}}
  \bibnamefont{and}
  \bibinfo{author}{\bibfnamefont{J.}~\bibnamefont{March-Russell}},
  \bibinfo{journal}{Phys. Lett.} \textbf{\bibinfo{volume}{B282}},
  \bibinfo{pages}{137} (\bibinfo{year}{1992}), \eprint{hep-th/9202003}.

\bibitem[{\citenamefont{Arkani-Hamed et~al.}(2001)\citenamefont{Arkani-Hamed,
  Cohen, and Georgi}}]{ArkaniHamed:2001ca}
\bibinfo{author}{\bibfnamefont{N.}~\bibnamefont{Arkani-Hamed}},
  \bibinfo{author}{\bibfnamefont{A.~G.} \bibnamefont{Cohen}}, \bibnamefont{and}
  \bibinfo{author}{\bibfnamefont{H.}~\bibnamefont{Georgi}},
  \bibinfo{journal}{Phys. Rev. Lett.} \textbf{\bibinfo{volume}{86}},
  \bibinfo{pages}{4757} (\bibinfo{year}{2001}), \eprint{hep-th/0104005}.

\bibitem[{\citenamefont{Hook}(2018)}]{Hook:2018jle}
\bibinfo{author}{\bibfnamefont{A.}~\bibnamefont{Hook}}, \bibinfo{journal}{Phys.
  Rev. Lett.} \textbf{\bibinfo{volume}{120}}, \bibinfo{pages}{261802}
  (\bibinfo{year}{2018}), \eprint{1802.10093}.

\end{thebibliography}

\end{document}